%% file: main.tex
\documentclass[runningheads]{llncs}
\usepackage{graphicx}
\usepackage{amsmath,amssymb} 
\usepackage{multirow}
\usepackage{diagbox}
\usepackage{gensymb}
\usepackage{subcaption}
\usepackage[export]{adjustbox}
\usepackage{xcolor}
\usepackage{stackengine}
\begin{document}
\title{DA-VSR: Domain Adaptable Volumetric Super-Resolution For Medical Images}
%
%
	\author{Cheng Peng\inst{1} \and
	S. Kevin Zhou\inst{2,3} \and
	Rama Chellappa\inst{1}}
	
	\institute{
	Johns Hopkins University, MD, USA \and 
	Medical Imaging, Robotics, and Analytic Computing Laboratory and Engineering (MIRACLE) Center, School of Biomedical Engineering \& Suzhou Institute for Advance Research, University of Science and Technology of China, Suzhou, China \and 
    Key Lab of Intelligent Information Processing of Chinese Academy of Sciences (CAS),
    Institute of Computing Technology, CAS, Beijing, China \\
    \email{cpeng26@jhu.edu, s.kevin.zhou@gmail.com, rchella4@jhu.edu}}
\maketitle              
\input{abstract}
\input{introduction}
\input{method}

\input{experiments}
\input{conclusion}
%
%
%
%
\bibliographystyle{splncs04}
\bibliography{bibliography}
%
\end{document}

%% file: abstract.tex
\begin{abstract}
Medical image super-resolution (SR) is an active research area that has many potential applications, including reducing scan time, bettering visual understanding, increasing robustness in downstream tasks, etc. However, applying deep-learning-based SR approaches for clinical applications often encounters issues of domain inconsistency, as the test data may be acquired by different machines or on different organs. In this work, we present a novel algorithm called domain adaptable volumetric super-resolution (DA-VSR) to better bridge the domain inconsistency gap. DA-VSR uses a unified feature extraction backbone and a series of network heads to improve image quality over different planes. Furthermore, DA-VSR leverages the in-plane and through-plane resolution differences on the test data to achieve a self-learned domain adaptation. As such, DA-VSR combines the advantages of a strong feature generator learned through supervised training and the ability to tune to the idiosyncrasies of the test volumes through unsupervised learning. Through experiments, we demonstrate that DA-VSR significantly improves super-resolution quality across numerous datasets of different domains, thereby taking a further step toward real clinical applications.
\end{abstract}

%% file: introduction.tex
\section{Introduction}

Medical imaging such as Magnetic Resonance Imaging (MRI) and Computed Tomography (CT) are crucial to clinical diagnosis. To facilitate faster and less costly acquisitions, it is routine to acquire a few high-resolution cross sectional images in CT/MRI, leading to a low through-plane resolution when the acquired images are organized into an anisotropic volume. The anisotropic volumes lead to difficulties in understanding the patient's anatomy both for physicians and automated algorithms \cite{peng2019deep,DBLP:conf/wacv/WangCWSG20}. One way to address this is through super-resolution (SR) algorithms \cite{zhou2021review,zhou2019handbook}, which upsample along the axis with a low resolution.
SR has witnessed great improvement in the image domain with Convolutional Neural Network (CNN)-based algorithms \cite{DBLP:journals/corr/DongLHT15,DBLP:journals/corr/KimLL15b,DBLP:conf/cvpr/KimLL16,DBLP:conf/cvpr/ZhangZGZ17}, where the formulation typically involves supervised learning between a low resolution (LR) image and its paired high resolution (HR) groundtruth. Various improvements have been made to reduce computation \cite{DBLP:journals/corr/DongLT16,shi2016realtime}, enhance feature extraction efficiency \cite{DBLP:conf/cvpr/Liu0T0W20,DBLP:conf/eccv/ZhangLLWZF18,DBLP:journals/corr/abs-1802-08797}, and improve robustness \cite{DBLP:conf/cvpr/ShocherCI18,DBLP:conf/cvpr/0008Z019}. 

Volumetric SR for medical images poses unique challenges. Firstly, the high dimensional and anisotropic nature of volumetric images lead to difficulties in computational cost and learning efficiency. While there exists many work on 2D medical image SR \cite{8736838,Park_2018,DBLP:journals/tip/ZhaoZZZ19,DBLP:journals/tip/CherukuriGSM20,DBLP:conf/icip/YuLSYWWCBH17,DBLP:journals/access/GeorgescuIV20}, few directly tackle 3D medical image SR due to high computational cost and limited sample size. Chen et al. \cite{DBLP:conf/miccai/ChenSCXZL18} apply a DenseNet-based CNN algorithms called mDCSRN on volumetric data with 3D kernels. Wang et al. \cite{DBLP:conf/wacv/WangCWSG20} ease the 3DCNN memory bottleneck by using a more efficiently implemented DenseNet with residual connections. These methods still require patch-by-patch inference on large-size volume, which can lead to undesirable border artifacts and inefficiency. Peng et al. \cite{DBLP:conf/cvpr/PengLLCZ20} propose SAINT, which can super-resolve volumetric images with multiple upsampling factors with a single network. Furthermore, it addresses the memory constraint at inference by performing an ensemble of 2D SR operations on through-plane images.  

Another challenge arises from the need for high reliability. Under a supervised learning framework, if a test image comes from a distribution not well represented in training, e.g. of a different body part or by a different machine, performance often degrades in unexpected ways. Therefore, semi-supervised or self-supervised SR methods provide distinct advantages if they can learn directly from test datasets. Zhao et al. \cite{DBLP:journals/tmi/ZhaoDPCRP21} propose SMORE, a self-supervised SR algorithm that leverages the high in-plane resolution to create LR-HR pairs for learning, and applies the learned model on lower through-plane resolution. Implicitly, SMORE assumes that the in-plane and through-plane images are from same or similar distributions, which may not be true for many cases.

To address the issue of robustly super-resolving volumetric medical images, we propose a novel algorithm named Domain Adaptable Volumetric Super-Resolution (DA-VSR). DA-VSR follows SAINT's thinking in addressing volumetric SR based on a series of slice-wise SR. DA-VSR uses a single feature extraction backbone and assigns small, task-specific network heads for upsampling and fusion. Inspired by SMORE, DA-VSR leverages the resolution differences across dimensions as a self-supervised signal for domain adaptation at \textit{test time}. Specifically, DA-VSR designs an additional self-supervised network head that can help align features on test images through in-plane super-resolution. As a result, DA-VSR enjoys the benefit of a strong feature backbone obtained by supervised training, and the ability to adapt to various distributions through unsupervised training. To summarize,
\begin{itemize}
  \item We design a slice-based volumetric SR network called DA-VSR. DA-VSR uses a Unified Feature Extraction (UFE) backbone and a series of lightweight network heads to perform super-resolution.
  \item We propose an in-plane SR head that propagates gradients to the UFE backbone both in training and testing. As such, DA-VSR is more robust, and can adapt its features to the test data distribution.
  \item We experiment with a diverse set of medical imaging data on different parts of the organ, and find large quantitative and visual improvement in SR quality on datasets out of the training distribution.
\end{itemize}

%% file: method.tex
\section{Domain Adaptable Volumetric Super-Resolution}
\begin{figure}[!htb]
    \centering
      \includegraphics[width=1\textwidth]{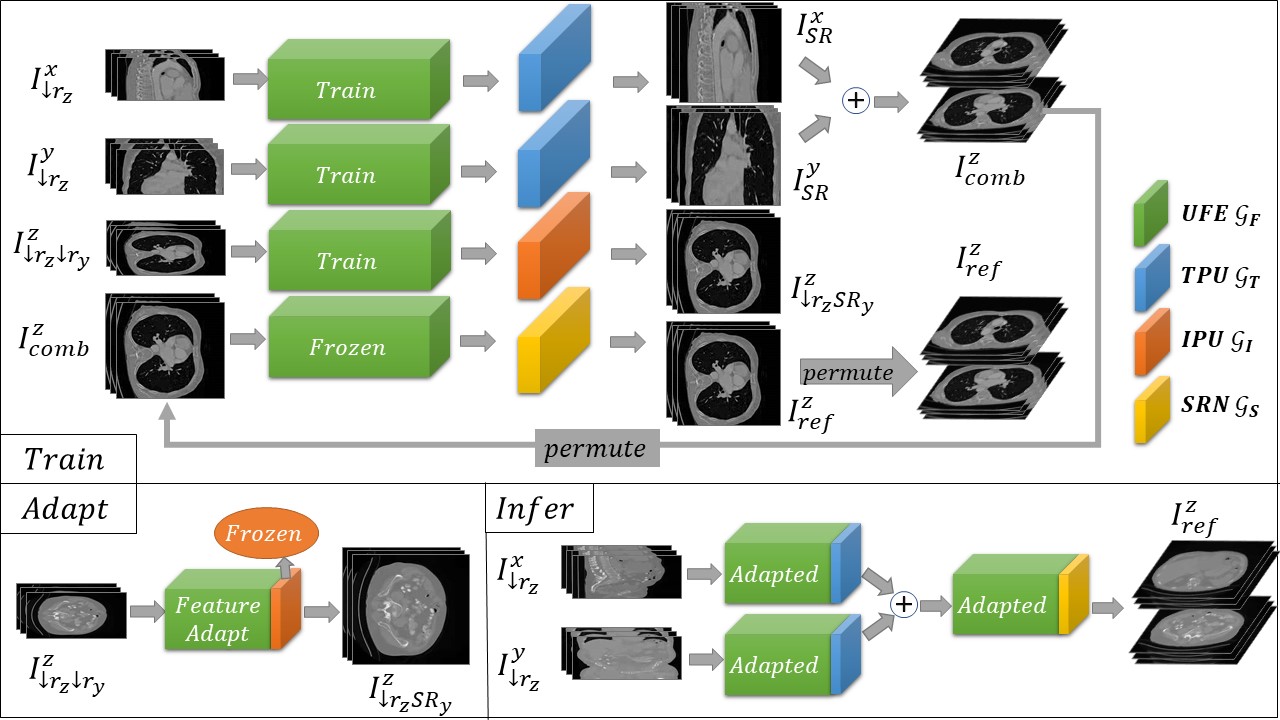}
    \caption{The overall pipeline of Domain Adaptable Volumetric Super-Resolution (DA-VSR). DA-VSR contains three stages. The network parameters are first trained in supervised setting. An additional adaptation stage is proposed to fit to test data. Finally inference is done through an adapted feature backbone and network heads. Networks of the same color share weights.}
    \label{fig:network}
\end{figure}

Consider $I(x,y,z) \in \mathbb{R}^{X\times Y\times Z}$ as a densely sampled volumetric medical image. Following the notations in \cite{DBLP:conf/cvpr/PengLLCZ20}, we refer to $x$, $y$, and $z$ as the sagittal, coronal, and axial axis, and $I^{x}(y,z)$, $I^{y}(x,z)$, and $I^{z}(x,y)$ as the sagittal, coronal, and axial slices. The task of super-resolution seeks to recover $I(x,y,z)$ from its partially observed, downsampled version $I_{\downarrow}(x,y,z)$. This work focuses on finding a transformation $\mathcal{F}\colon{\mathbb{R}^{X\times Y\times \frac{Z}{r_z}}}\to{\mathbb{R}^{X\times Y\times Z}}$ that super-resolves an axially sparse volume $I_{\downarrow r_z}(x,y,z)$ to $I(x,y,z)$, where $r_z$ is the sparsity factor in the axial axis. 
The super-resolving function $\mathcal{F}$ is most popularly approximated through a CNN $\mathcal{F}_{\theta_{\mathcal{S}}}$, where $\theta_{\mathcal{S}}$ denotes the network parameters learned from a training set $\mathcal{S}$ that contains LR-HR pairs $\{I^{\mathcal{S}}_{\downarrow r_z}(x,y,z),I^{\mathcal{S}}(x,y,z)\}$ for supervised learning. While $\mathcal{F}_{\theta_{\mathcal{S}}}$ may be near-optimal for $\mathcal{S}$, its performance degrades when used on a test set $\mathcal{T}$ from a different distribution. Hence, we would like to approximate a better $\mathcal{F}_{\theta_{\mathcal{T}}}$ based on $\theta_{\mathcal{S}}$ and $I^{\mathcal{T}}_{\downarrow r_z}(x,y,z)$. As shown in Fig. \ref{fig:network}, DA-VSR consists of a supervised training stage that yields an $\mathcal{F}_{\theta_{\mathcal{S}}}$, and a self-supervised adaptation stage that adjusts its network parameters to fit with $\mathcal{T}$.

\subsection{Network Structure}
DA-VSR consists of four components: a Unified Feature Extraction (UFE) backbone and three lightweight network heads called Through-Plane Upsampler (TPU), In-Plane Upsampler (IPU), and Slice Refinement Net (SRN), which are denoted as $\mathcal{G}_{F}$, $\mathcal{G}_{T}$, $\mathcal{G}_{I}$, and $\mathcal{G}_{S}$, respectively. 

DA-VSR upsamples $I_{\downarrow}(x,y,z)$ by first performing a sequence of 2D SR on the low resolution through-plane slices $I^{x}_{\downarrow r_z}(y,z)$ and $I^{y}_{\downarrow r_z}(x,z)$ using UFE $\mathcal{G}_{F}$ and TPU $\mathcal{G}_{T}$, which can be described as follows:
\begin{align} \label{eq:TPU}
    I^{x}_{SR}(y,z)=\mathcal{G}_{T}\circ\mathcal{G}_{F}(I^{x}_{\downarrow r_z}(y,z)),~I^{y}_{SR}(x,z)=\mathcal{G}_{T}\circ\mathcal{G}_{F}(I^{y}_{\downarrow r_z}(x,z)).
\end{align}
The loss for training $\mathcal{G}_{T}\circ\mathcal{G}_{F}$ is defined as:
\begin{align}\label{eq:loss_tpu}
    \mathcal{L}_{tpu} = \lVert I^{x}_{SR} - I^{x}_{gt} \rVert_1 + \lVert I^{y}_{SR} - I^{y}_{gt} \rVert_1.
\end{align}

DA-VSR also performs an in-plane slice upsampling by first downsampling the high resolution  $I^{z}_{\downarrow r_z}(x,y)$ to $I^{z}_{\downarrow r_z\downarrow r_y}(x,y)$ over the $y$ axis, where $r_y = r_z$. $I^{z}_{\downarrow r_z\downarrow r_x}(x,y)$ can be similarly generated with an additional permutation to ensure agreement in input/output dimensions. This self-learning process that uses UFE $\mathcal{G}_{F}$ and IPU $\mathcal{G}_{I}$ is introduced during training as:
\begin{align} \label{eq:IPU}
    I^{z}_{\downarrow r_zSR_x}(x,y)=\mathcal{G}_{I}\circ\mathcal{G}_{F}(I^{z}_{\downarrow r_z\downarrow r_x}(x,y)),~I^{z}_{\downarrow r_zSR_y}(x,y)=\mathcal{G}_{I}\circ\mathcal{G}_{F}(I^{z}_{\downarrow r_z\downarrow r_y}(x,y)).
\end{align}

The loss formulation for training $\mathcal{G}_{I}\circ\mathcal{G}_{F}$ is:
\begin{align}\label{eq:loss_ipu}
    \mathcal{L}_{ipu} = \lVert I^{z}_{\downarrow r_zSR_x} - I^{z}_{\downarrow r_z} \rVert_1 +\lVert I^{z}_{\downarrow r_zSR_y} - I^{z}_{\downarrow r_z} \rVert_1.
\end{align}

During training, the overall loss is defined as $\mathcal{L}_{main} = \lambda_{tpu}*\mathcal{L}_{tpu} + \lambda_{ipu}*\mathcal{L}_{ipu}$, where $\lambda_{tpu}$ and $\lambda_{ipu}$ are selected as 2 and 1 empirically. After $\mathcal{G}_{F}$, $\mathcal{G}_{T}$, and $\mathcal{G}_{I}$ are trained to convergence, we \underline{freeze} their network parameters to obtain $\mathcal{G}^{fro}_{F}$, $\mathcal{G}^{fro}_{T}$, and $\mathcal{G}^{fro}_{I}$, respectively. The super-resolved slices $I^{x}_{SR}$ and $I^{y}_{SR}$ are reformatted into volumes and averaged to yield a single volume $I_{comb}(x,y,z) = \frac{1}{2}( I^{x}_{SR}(x,y,z) + I^{y}_{SR}(x,y,z))$. We then feed the axial slices $I^{z}_{comb}(x,y)$ into a frozen UFE $\mathcal{G}^{fro}_{F}$ and a SRN $\mathcal{G}_{S}$. As such, we reuse the already well-trained feature extraction from a deep $\mathcal{G}_{F}$ and a lightweight $\mathcal{G}_{S}$ to perform axial refinement in $I^{z}_{comb}(x,y)$. Note that this is different from SAINT, which trains an independent, relatively shallow network from scratch to perform axial refinement. The forward process and training loss are:
\begin{align} \label{eq:SRN}
    I^{z}_{ref}(x,y)=\mathcal{G}^{fro}_{S}\circ\mathcal{G}_{F}(I^{z}_{comb}(x,y)), \mathcal{L}_{ref} = \lVert I^{z}_{ref} - I^{z}_{gt} \rVert_1.
\end{align}

\subsection{Self-Supervised Adaptation}

After all networks are trained to convergence with dataset $\mathcal{S}$, DA-VSR uses a simple yet effective adaptation stage before inference on test set $\mathcal{T}$, as inspired by SMORE\cite{DBLP:journals/tmi/ZhaoDPCRP21}. We seek to update network parameters in $\mathcal{G}_{F}$ to fit to $\mathcal{T}$ through the common task of in-plane super-resolution. In particular, we \underline{freeze $\mathcal{G}_{I}$} during adaptation and only allow backpropogating gradients to modify parameters in $\mathcal{G}_{F}$ to form $\mathcal{G}^{adp}_{F}$. This process is described as:
\begin{align} \label{eq:IPU_fro}
    I^{z}_{\downarrow r_zSR_x}(x,y)=\mathcal{G}^{adp}_{I}\circ\mathcal{G}^{fro}_{F}(I^{z}_{\downarrow r_z\downarrow r_x}(x,y)),~I^{z}_{\downarrow r_zSR_y}(x,y)=\mathcal{G}^{adp}_{I}\circ\mathcal{G}^{fro}_{F}(I^{z}_{\downarrow r_z\downarrow r_y}(x,y)),
\end{align}
where $\mathcal{G}^{fro}_{I}$ denotes a frozen, pretrained $\mathcal{G}_{I}$. The loss formulation is similar to Eq. (\ref{eq:loss_ipu}). We find that this approach effectively prevents overfitting of adaptation to the test data $\mathcal{T}$. If $\mathcal{G}_{I}$ is not frozen, the composition of $\mathcal{G}_{I}\circ\mathcal{G}_{G}$ effectively becomes a SMORE setup after tuning. With $\mathcal{G}_{T}$ and $\mathcal{G}_{I}$ both learned through the training data, $\mathcal{G}_{I}$ serves as proxy that constrains features generated by $\mathcal{G}_{F}$ to not veer far from the effective range of $\mathcal{G}_{T}$.
As shown in Fig. \ref{fig:network}, after self-supervised adaptation, inference can be done straigtforwardly. The LR through-plane slices $I^{x}_{\downarrow r_z}(y,z)$ and $I^{y}_{\downarrow r_z}(x,z)$ are upsampled by $\mathcal{G}^{adp}_{T}\circ\mathcal{G}^{fro}_{F}$, combined together as a volume through averaging, and fed to $\mathcal{G}^{adp}_{S}\circ\mathcal{G}^{fro}_{F}$ as a series of axial slices. The refined slices $I^{z}_{ref}$ are formatted to form the final upsampled volume.

%% file: experiments.tex
\section{Experiments}

\subsection{Implementation Details}
To ensure a fair comparison, we implement all compared models to have similar number of parameters, as shown in Table \ref{tab:ablation_table} and Table \ref{tab:sota_table}. For DA-VSR, we use Residual Dense Blocks (RDBs)\cite{DBLP:journals/corr/abs-1802-08797} as the building block for UFE. Specifically, we use six RDBs, each of which has eight convolution layers and a growth rate of thirty-two. For TPU, IPU and SRN, we use a lightweight design of three convolution layers. TPU and IPU use additional pixel shuffling layers to upsample the spatial dimension on the axial axis. Following \cite{DBLP:conf/cvpr/PengLLCZ20}, the LR input image is formatted as three consecutive slices.

 SAINT\cite{DBLP:conf/cvpr/PengLLCZ20} is similarly implemented with a RDB-based network. 3DRDN is implemented with two RDBs, each containing eight convolutional layers and with a growth rate of thrity-two. 3DRCAN is implemented with three residual groups, each containing three residual blocks and with a feature dimension of sixty-four. 

\subsection{Dataset}
For training and validation, we use 890 CT scans from the publicly available LIDC-IDRI\cite{armato2011lung} dataset, which are taken on lungs. We use 810 volumes for training, 30 volumes for validation, and 50 volumes for testing. Additionally, we use a slew of CT datasets that are acquired for different organs, including liver \cite{simpson2019large}, colon \cite{simpson2019large}, and kidney \cite{1904.00445} for testing. For consistency and following previous literature \cite{DBLP:conf/wacv/WangCWSG20,DBLP:conf/miccai/ChenSCXZL18}, all volumes are selected with slice thickness between 1mm to 3.5mm, and interpolated to 2.5mm. We then discard volumes with less than 128 axial slices. After pre-processing, we obtain 120 liver volumes, 30 colon volumes, and 59 kidney volumes for testing. Due to large memory cost at inference time for 3D-kernel baselines, in-plane pixel resolution is downsampled from $512\times512$ to $256\times256$. For training, 3D-kernel baselines use patches of size $64\times64\times Z$. 

\input{images/vis1}
\subsection{Ablation Study}
We examine the effectiveness of through-plane SR in DA-VSR as compared to other implementations. Specifically, we compare the full DA-VSR to:
\begin{itemize}
    \item $\textrm{DA-VSR}_{SMORE}$: DA-VSR that only uses test data to train, implemented similar to SMORE \cite{DBLP:journals/tmi/ZhaoDPCRP21}.
    \item $\textrm{DA-VSR}_{SAINT}$: DA-VSR without a self-supervised $\mathcal{G}_{I}$, similar to SAINT\cite{DBLP:conf/cvpr/PengLLCZ20}.
    \item $\textrm{DA-VSR}^{NA}$: DA-VSR without a self-supervised adaptation stage.
    \item $\textrm{DA-VSR}^{A}_{nofro}$: DA-VSR with a self-supervised adaptation stage, where $\mathcal{G}_{I}$ is not frozen during adaptation.
\end{itemize}

\input{tables/quantitative}

The performances are summarized in Table \ref{tab:ablation_table}. Improvements can be seen from $\textrm{DA-VSR}^{NA}$ to the full DA-VSR when self-supervised adaptation is applied. We also observe that if $\mathcal{G}_{I}$ is not frozen during adaptation, performance is severely degraded after adaptation. Note for adaptation, $\textrm{DA-VSR}^{A}_{nofro}$ and DA-VSR are both trained to convergence, which typically takes five to ten epochs. In comparison, adaptation on DA-VSR with a frozen $\mathcal{G}^{fro}_{I}$ is stable and performance does not degrade even if trained over many epochs. This can be helpful as no scheme for early stopping is required. We also observe that quantitatively $\textrm{DA-VSR}^{NA}$ performs slightly better over unseen datasets than $\textrm{DA-VSR}_{SAINT}$, despite slightly worse performance over lung. This may be attributed to DA-VSR's self-supervised in-plane upsampling process, as it forces the network observe a more diverse data distribution. Finally, $\textrm{DA-VSR}_{SMORE}$ performs well over unseen datasets and in some instances is nearly on par with supervised methods, e.g. over the Liver dataset. The performance of $\textrm{DA-VSR}_{SMORE}$ fluctuates depending on (1) how similar in-plane and through-plane statistics are, and (2) the sample size of the test dataset. In comparison, DA-VSR is less reliant on these factors, as its feature extractor is trained with supervision and on a large dataset.

\subsection{Quantitative Evaluation}
As shown in Table \ref{tab:sota_table}, we compare the full DA-VSR pipeline against other state-of-the-art SR implementations: a 3D-kernel variant of RDN \cite{DBLP:conf/cvpr/ZhangZGZ17}, a 3D-kernel variant of RCAN \cite{DBLP:conf/eccv/ZhangLLWZF18}, and SAINT \cite{DBLP:conf/cvpr/PengLLCZ20}. We find that 3D RDN and RCAN is not as efficient as a slice-based volumetric SR approach like SAINT under similar network size. As an overall pipeline, DA-VSR performs slightly better than SAINT on unseen dataset, and significantly better when adaptation is applied. We find that by using a unified feature backbone to generate features, DA-VSR's slice refinement stage converges much faster than SAINT's despite using less parameters; please refer to the supplemental material for illustration. 

\input{tables/sota}
While metrics like PSNR and SSIM are useful to understand performance in aggregate, they can often be too coarse. Domain drift does not happen uniformly on a CT image. Some patches do not suffer as much since similar patches can be observed in the training set, leading to similar overall PSNR metrics; however, some patches suffer heavily due to lack of observations. We provide visualization on SR results, as shown in Fig. \ref{fig:ablation_visual}, to better show where improvements are most often observed. We observe that supervised techniques without adaptation can lead to significant overfitting issues over unseen test sets and create unfaithful details, as seen in Case 2 of Fig. \ref{fig:ablation_visual} and indicated by the orange arrows. Compared to supervised methods, SMORE produces results that are smoother and more similar to the groundtruth if those patterns are seen in axial slices. For organs that exhibit different patterns between axial and other axes, such as the sagittal spinal structure, SMORE can generate unreliable or overly smooth patterns, as shown in Case 1 of Fig. \ref{fig:ablation_visual}. In comparison, since DA-VSR goes through supervised training on a lung dataset, which contains both LR and HR spine patterns, we observe that it performs much better than SMORE even with adaptation. Interestingly, we also observe improvements over the few unseen cases where a region of lung is included, e.g. in Case 3, by using our proposed adaptation stage. Despite being trained on the lung dataset, other supervised methods still experience local discontinuity over small scale bone structures. In this case, SMORE generates smoother but less structurally reliable details. DA-VSR takes the advantages of both approaches and generate smoother and more reliable details under this challenging case. As no two individuals are the same, DA-VSR's ability to reduce minor distribution differences can be valuable in real SR applications. 

%% file: images/vis1.tex
\begin{figure}[!t]
	\scriptsize
    \begin{tabular}[b]{cc}
        \begin{subfigure}[b]{0.38\linewidth}
            \includegraphics[width=\linewidth,height=0.6\linewidth,cframe=red]{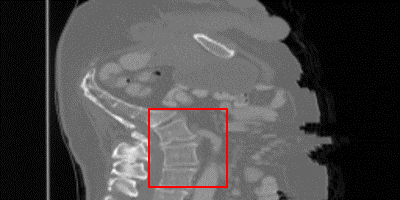}
        \caption*{\stackanchor{Case 1}{}}
        \end{subfigure}
        &
        \begin{tabular}[b]{c c c c}
            \begin{subfigure}[b]{0.15\linewidth}
                \setlength{\abovecaptionskip}{3pt}
                \includegraphics[width=\linewidth,height=0.5\linewidth,cframe=red]{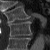}
                \caption*{\stackanchor{$\textrm{HR}_{}$}{PSNR/SSIM}}
            \end{subfigure}
            &
            \begin{subfigure}[b]{0.15\linewidth}
                \setlength{\abovecaptionskip}{3pt}
                \includegraphics[width=\linewidth,height=0.5\linewidth,cframe=red]{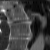}
                \caption*{\stackanchor{$\textrm{DA-VSR}_{}$}{\bf{37.56/0.971}}}
            \end{subfigure}
            &
            \begin{subfigure}[b]{0.15\linewidth}
                \setlength{\abovecaptionskip}{3pt}
                \includegraphics[width=\linewidth,height=0.5\linewidth,cframe=red]{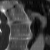}
                \caption*{\stackanchor{$\textrm{DA-VSR}_{NA}$}{37.02/0.968}}
            \end{subfigure}
            &
            \begin{subfigure}[b]{0.15\linewidth}
                \setlength{\abovecaptionskip}{3pt}
                \includegraphics[width=\linewidth,height=0.5\linewidth,cframe=red]{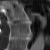}
                \caption*{\stackanchor{$\textrm{SAINT}_{}$}{36.89/0.966}}
            \end{subfigure}\\
            \begin{subfigure}[b]{0.15\linewidth}
                \setlength{\abovecaptionskip}{3pt}
                \includegraphics[width=\linewidth,height=0.5\linewidth,cframe=red]{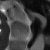}
                \caption*{\stackanchor{SMORE}{35.99/0.960}}
            \end{subfigure}
            &
            \begin{subfigure}[b]{0.15\linewidth}
                \setlength{\abovecaptionskip}{3pt}
                \includegraphics[width=\linewidth,height=0.5\linewidth,cframe=red]{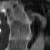}
                \caption*{\stackanchor{3DRDN}{36.40/0.962}}
            \end{subfigure}
            &
            \begin{subfigure}[b]{0.15\linewidth}
                \setlength{\abovecaptionskip}{3pt}
                \includegraphics[width=\linewidth,height=0.5\linewidth,cframe=red]{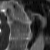}
                \caption*{\stackanchor{3DRCAN}{36.54/0.962}}
            \end{subfigure}
            &
            \begin{subfigure}[b]{0.15\linewidth}
                \setlength{\abovecaptionskip}{3pt}
                \includegraphics[width=\linewidth,height=0.5\linewidth,cframe=red]{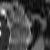}
                \caption*{\stackanchor{Bicubic}{32.49/0.915}}
            \end{subfigure}
        \end{tabular}
    \end{tabular}\\
    \begin{tabular}[b]{cc}
        \begin{subfigure}[b]{0.38\linewidth}
            \includegraphics[width=\linewidth,height=0.6\linewidth,cframe=red]{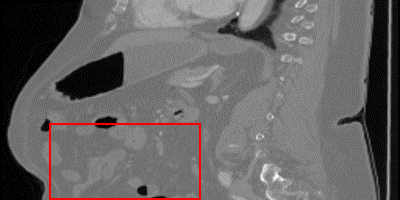}
        \caption*{\stackanchor{Case 2}{}}
        \end{subfigure}
        &
        \begin{tabular}[b]{c c c c}
            \begin{subfigure}[b]{0.15\linewidth}
                \setlength{\abovecaptionskip}{3pt}
                \includegraphics[width=\linewidth,height=0.5\linewidth,cframe=red]{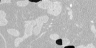}
                \caption*{\stackanchor{$\textrm{HR}_{}$}{PSNR/SSIM}}
            \end{subfigure}
            &
            \begin{subfigure}[b]{0.15\linewidth}
                \setlength{\abovecaptionskip}{3pt}
                \includegraphics[width=\linewidth,height=0.5\linewidth,cframe=red]{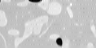}
                \caption*{\stackanchor{$\textrm{DA-VSR}_{}$}{\bf{34.91/0.980}}}
            \end{subfigure}
            &
            \begin{subfigure}[b]{0.15\linewidth}
                \setlength{\abovecaptionskip}{3pt}
                \includegraphics[width=\linewidth,height=0.5\linewidth,cframe=red]{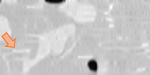}
                \caption*{\stackanchor{$\textrm{DA-VSR}_{NA}$}{34.78/0.976}}
            \end{subfigure}
            &
            \begin{subfigure}[b]{0.15\linewidth}
                \setlength{\abovecaptionskip}{3pt}
                \includegraphics[width=\linewidth,height=0.5\linewidth,cframe=red]{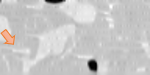}
                \caption*{\stackanchor{$\textrm{SAINT}_{}$}{34.84/0.977}}
            \end{subfigure}\\
            \begin{subfigure}[b]{0.15\linewidth}
                \setlength{\abovecaptionskip}{3pt}
                \includegraphics[width=\linewidth,height=0.5\linewidth,cframe=red]{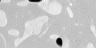}
                \caption*{\stackanchor{SMORE}{34.65/0.978}}
            \end{subfigure}
            &
            \begin{subfigure}[b]{0.15\linewidth}
                \setlength{\abovecaptionskip}{3pt}
                \includegraphics[width=\linewidth,height=0.5\linewidth,cframe=red]{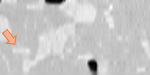}
                \caption*{\stackanchor{3DRDN}{34.47/0.966}}
            \end{subfigure}
            &
            \begin{subfigure}[b]{0.15\linewidth}
                \setlength{\abovecaptionskip}{3pt}
                \includegraphics[width=\linewidth,height=0.5\linewidth,cframe=red]{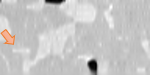}
                \caption*{\stackanchor{3DRCAN}{34.60/0.965}}
            \end{subfigure}
            &
            \begin{subfigure}[b]{0.15\linewidth}
                \setlength{\abovecaptionskip}{3pt}
                \includegraphics[width=\linewidth,height=0.5\linewidth,cframe=red]{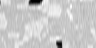}
                \caption*{\stackanchor{Bicubic}{31.60/0.926}}
            \end{subfigure}
        \end{tabular}
    \end{tabular}    \\
    \begin{tabular}[b]{cc}
        \begin{subfigure}[b]{0.38\linewidth}
            \includegraphics[width=\linewidth,height=0.6\linewidth,cframe=red]{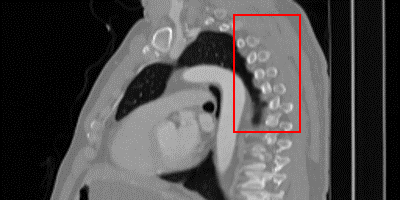}
        \caption*{\stackanchor{Case 3}{}}
        \end{subfigure}
        &
        \begin{tabular}[b]{c c c c}
            \begin{subfigure}[b]{0.15\linewidth}
                \setlength{\abovecaptionskip}{3pt}
                \includegraphics[width=\linewidth,height=0.5\linewidth,cframe=red]{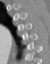}
                \caption*{\stackanchor{$\textrm{HR}_{}$}{PSNR/SSIM}}
            \end{subfigure}
            &
            \begin{subfigure}[b]{0.15\linewidth}
                \setlength{\abovecaptionskip}{3pt}
                \includegraphics[width=\linewidth,height=0.5\linewidth,cframe=red]{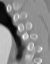}
                \caption*{\stackanchor{$\textrm{DA-VSR}_{}$}{\bf{37.83/0.978}}}
            \end{subfigure}
            &
            \begin{subfigure}[b]{0.15\linewidth}
                \setlength{\abovecaptionskip}{3pt}
                \includegraphics[width=\linewidth,height=0.5\linewidth,cframe=red]{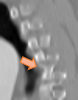}
                \caption*{\stackanchor{$\textrm{DA-VSR}_{NA}$}{36.68/0.969}}
            \end{subfigure}
            &
            \begin{subfigure}[b]{0.15\linewidth}
                \setlength{\abovecaptionskip}{3pt}
                \includegraphics[width=\linewidth,height=0.5\linewidth,cframe=red]{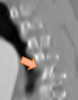}
                \caption*{\stackanchor{$\textrm{SAINT}_{}$}{36.77/0.969}}
            \end{subfigure}\\
            \begin{subfigure}[b]{0.15\linewidth}
                \setlength{\abovecaptionskip}{3pt}
                \includegraphics[width=\linewidth,height=0.5\linewidth,cframe=red]{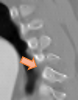}
                \caption*{\stackanchor{SMORE}{35.08/0.954}}
            \end{subfigure}
            &
            \begin{subfigure}[b]{0.15\linewidth}
                \setlength{\abovecaptionskip}{3pt}
                \includegraphics[width=\linewidth,height=0.5\linewidth,cframe=red]{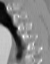}
                \caption*{\stackanchor{3DRDN}{36.50/0.968}}
            \end{subfigure}
            &
            \begin{subfigure}[b]{0.15\linewidth}
                \setlength{\abovecaptionskip}{3pt}
                \includegraphics[width=\linewidth,height=0.5\linewidth,cframe=red]{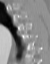}
                \caption*{\stackanchor{3DRCAN}{36.42/0.967}}
            \end{subfigure}
            &
            \begin{subfigure}[b]{0.15\linewidth}
                \setlength{\abovecaptionskip}{3pt}
                \includegraphics[width=\linewidth,height=0.5\linewidth,cframe=red]{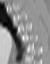}
                \caption*{\stackanchor{Bicubic}{31.62/0.927}}
            \end{subfigure}
        \end{tabular}
    \end{tabular}       
    \caption{Visual comparisons of DA-VSR and other state-of-the-art implementations from the sagittal plane, highlight regions are contrast-adjusted. Case 1 is from the Colon dataset, Case 2 and 3 are from the Kidney dataset\cite{1904.00445}. In particular, Case 3 is a lung region cropped from a kidney-containing image. Please refer to the supplemental material for more visual comparisons.} 
    \label{fig:ablation_visual}
    \vspace{-1em}
\end{figure}

%% file: tables/quantitative.tex
\begin{table}[!htb]
\caption{Quantitative ablation study of \underline{through-plane} upsampling for DA-VSR in terms of PSNR and SSIM, measured based on $I^{x}_{SR}(y,z)$. The best results are in {\bf bold}, and the second best results are \underline{underlined}. All baselines achieved similar performance compared to the original papers.}
\label{tab:ablation_table}
\centering 
\begin{tabular}{|l|l|r|r|rrr|}
\hline
 Scale & Method & Param & Lung &Liver & Colon & Kidney \\
\hline
\multirow{5}{*}{X4}& $\textrm{DA-VSR}_{SMORE}$ & 2.8M& 38.25/0.971\textbf{} & 39.05/0.981 &  40.00/0.986 & 37.18/0.975\\
&$\textrm{DA-VSR}_{SAINT}$ &2.8M& \bf{39.77/0.976} & 39.11/0.980& 40.32/0.987  & 37.53/0.975\\
&$\textrm{DA-VSR}^{NA}$ & 2.9M&\underline{39.67/0.976} & \underline{39.29/0.981}& \underline{40.43/0.987} &  \underline{37.60/0.976} \\
&$\textrm{DA-VSR}^{A}_{nofro}$ & 2.9M& N/A & 38.94/0.981 & 39.89/0.983 & 36.91/0.974\\
&DA-VSR &2.9M&N/A & \bf{39.51/0.982}& \bf{40.60/0.988} &  \bf{38.07/0.978} \\
\hline
\end{tabular}
\end{table}

%% file: tables/sota.tex
\begin{table}[!htb]
\caption{Quantitative \underline{volume-wise} evaluation of DA-VSR against SoTA SR implementations in terms of PSNR and SSIM. The best results are in {\bf bold}, and the second best results are \underline{underlined}. All baselines achieved similar performance compared to the original papers.}
\label{tab:sota_table}
\centering 
\begin{tabular}{|l|l|r|r|rrr|}
\hline
 Scale & Method & Param & Lung &Liver & Colon & Kidney \\
\hline
\multirow{5}{*}{X4}& Bicubic  &N/A & 33.72/0.941& 34.56/0.955 & 35.23/0.964 & 33.37/0.948 \\
&3D RCAN &2.9M& 39.30/0.975 & 38.95/0.979 & 40.11/0.986 & 37.41/0.975\\
&3D RDN & 2.9M& 39.39/0.976& 39.04/0.980 & 40.22/0.986 & 37.46/0.975 \\
&SAINT &2.9M& \bf{40.01/0.977} & 39.30/0.981 &  40.51/0.987 & 37.70/0.976\\
&$\textrm{DA-VSR}^{NA}$ &3.0M& \underline{39.90/0.977} & \underline{39.48/0.981} & \underline{40.68/0.988} &  \underline{37.82/0.977}\\
&DA-VSR &3.0M&N/A & \bf{39.74/0.983} & \bf{40.83/0.988} &  \bf{38.28/0.979} \\

\hline
\multirow{5}{*}{X6}&Bicubic  & N/A& 31.47/0.913 & 32.55/0.935 & 32.98/0.943 & 31.31/0.924 \\
&3D RCAN &3.0M& 36.56/0.962 & 36.62/0.970 & 37.19/0.975 & 35.04/0.962\\
&3D RDN &3.3M& 36.73/0.963 & 36.67/0.970 &37.23/0.976 & 35.10/0.963\\
&SAINT & 2.9M& \bf{37.23/0.966} &36.76/0.971 &37.37/0.977 & 35.26/0.964\\ 
&$\textrm{DA-VSR}^{NA}$ & 3.0M&\underline{37.14/0.965}& \underline{36.86/0.972} & \underline{37.44/0.977} & \underline{35.31/0.964}\\
&DA-VSR&3.0M& N/A  &\bf{37.18/0.973} & \bf{37.78/0.979} &\bf{35.55/0.966}\\

\hline
\end{tabular}
\end{table}

%% file: conclusion.tex
\section{Conclusion}
We propose a Domain-Adpatable Volumetric Super-Resolution (DA-VSR). Inspired by SAINT \cite{DBLP:conf/cvpr/PengLLCZ20} and SMORE \cite{DBLP:journals/tmi/ZhaoDPCRP21}, DA-VSR leverages the advantages in supervised and self-supervised learning. Specifically, DA-VSR uses supervised training to learn a strong feature generator with various task-specific network heads, and a self-supervised domain adaptation stage to better fit to unseen test sets. We carefully evaluate our approach between training and testing CT datasets that are acquired on different organs. We find that DA-VSR produce consistent improvements in quantitative measurements and visual quality. Our approach is conceptually straightforward and can be implemented with different network structures. Future work includes investigating the effect of anisotropic resolution to self-supervised adaptation and the effect of DA-VSR over other types of domain gap, e.g. cross machine, cross modality, etc. 